\documentclass[12pt]{article}

\usepackage{epsfig}
\usepackage{amssymb,amsmath}
\usepackage{setspace}

\newcommand{\bea}{\begin{eqnarray}}
\newcommand{\eea}{\end{eqnarray}}

 \makeatletter
\def\alt{\mathrel{\mathpalette\gl@align<}}
\def\agt{\mathrel{\mathpalette\gl@align>}}
\def\gl@align#1#2{\lower.6ex\vbox{\baselineskip\z@skip\lineskip\z@
\ialign{$\m@th#1\hfil##\hfil$\crcr#2\crcr\sim\crcr}}} \makeatother

\begin{document}
%
\vspace*{1.0cm}

\begin{center}
\baselineskip 20pt {\Large\bf
125 GeV Higgs Boson From Gauge-Higgs Unification: \\
A Snowmass white paper
}
\vspace{1cm}

{\large
Ilia Gogoladze$^{a}$\footnote{ E-mail:ilia@bartol.udel.edu},
Nobuchika Okada$^{b}$\footnote{E-mail:okadan@ua.edu}
and Qaisar Shafi$^{a}$\footnote{ E-mail:shafi@bartol.udel.edu}
} \vspace{.5cm}

{\baselineskip 20pt \it
$^a$Bartol Research Institute, Department of Physics and Astronomy, \\
University of Delaware, Newark, DE 19716, USA \\
\vspace{2mm}
$^b$Department of Physics and Astronomy,\\
University of Alabama, Tuscaloosa, AL 35487, USA
}
\vspace{.5cm}

\vspace{1.5cm} {\bf Abstract}
\end{center}

In certain five dimensional gauge theories compactified
 on the orbifold $S^1/Z_2$
 the Standard Model Higgs doublet is identified
 with the zero mode of the fifth component of the gauge field.
This gauge-Higgs unification scenario is realized
 at high energies, and the Standard Model as an effective theory
 below the compactification scale satisfies the boundary
 condition that the Higgs quartic coupling vanishes
 at the compactification scale (gauge-Higgs condition).
This is because at energies above the compactification scale,
 the five dimensional gauge invariance is restored and
 the Higgs potential vanishes as a consequence.
We consider scenario where top quark Yukawa and weak gauge 
 coupling unification can be realized and 
 identify the compactification scale as one 
 at which  this two coupling couplings have the same value. 
Taking into account the experimental uncertainties 
 in measurements of the top quark mass and the QCD coupling constant, 
 the Higgs mass prediction of 119-126 GeV 
 from the gauge-Higgs unification scenario 
 is consistent with the experimentally measured 
 value of 125-126 GeV.
More precise measurements of the top quark mass 
 and the QCD coupling constant are crucial 
 to reduce the interval of the Higgs mass prediction and 
 thereby test the feasibility of the gauge-Higgs unification
 scenario. 

\thispagestyle{empty}

\newpage

\addtocounter{page}{-1}

\baselineskip 18pt

A major goal of the physics program at the Large Hadron Collider
 is to confirm the origin of the electroweak symmetry
 breaking and the mechanism of particle mass generation.
Both the ATLAS~\cite{ATLAS} and CMS~\cite{CMS} experiments
 have discovered, it seems, the Higgs(-like) boson with a mass
 of around 125 GeV through a variety of Higgs boson decay modes.
With the discovery of the Higgs boson, the experimental tests
 of the Higgs sector of the Standard Model (SM) have just started.

In the SM, the Higgs boson mass is determined
 by the quartic coupling of the Higgs doublet, so that
 the measured Higgs boson mass provides us information
 of the Higgs quartic coupling constant at the electroweak scale.
This information has a great impact on the SM
 as well as new physics beyond the SM
 which make predictions of the Higgs boson mass.
For example, a mass of around 125 GeV is very interesting
 from the viewpoint of the vacuum stability bound
 on the Higgs boson mass~\cite{Buttazzo:2013uya}.
If the Higgs quartic coupling corresponding to 125 GeV is
 extrapolated to high energies according to the renormalization group
 equations of the SM, the coupling becomes
 negative at an intermediate scale of around $10^9-10^{11}$ GeV,
 depending on experimental uncertainties in top quark mass
 and the QCD coupling constant.
Thus, the electroweak vacuum is not stable.

This vacuum instability may indicate that
 new physics beyond the SM appears
 at or below the intermediate scale,
 which modifies the renormalization group equations
 in the SM and prevents the running
 Higgs quartic coupling from becoming negative
 up to the Planck mass~\cite{Gogoladze:2008gf}.
In the following, we consider a five dimensional gauge-Higgs unification
 scenario~\cite{GHU} (see also \cite{Antoniadis}), 
 compactified on the orbifold $S^1/Z_2$.
Through compactification and non-trivial parity assignments
 for the five dimensional gauge fields, the original gauge group
 is broken to the SM gauge group,
 with the Higgs doublet identified with the fifth component
 of the gauge field.
In the four dimensional effective theory,
 the five dimensional gauge invariance requires
 a vanishing Higgs potential at tree level.
The Higgs potential of the SM is generated
 below the compactification scale through quantum effects
 associated with the Kaluza-Klein modes of bulk fields.

Below the compactification scale, all Kaluza-Klein modes
 are decoupled and the SM is realized
 as an effective theory at low energies.
Although some effort is needed to construct a realistic scenario
 (see for example \cite{GHU-Models}),
 the most important property of the gauge-Higgs
 unification scenario is encoded in the Higgs quartic coupling.
Namely, the Higgs quartic coupling should vanish
 once the higher dimensional gauge invariance gets restored.
In fact, it has been explicitely shown~\cite{HMOY}
 that the effective Higgs quartic coupling calculated
 in a given gauge-Higgs unification model coincides with
 the one generated through the renormalization group equations
 of the SM with a boundary condition that
 the Higgs quartic coupling vanishes at the compactification scale.
Applying this gauge-Higgs condition, we can predict the Higgs boson mass
 as a function of the compactification scale~\cite{GHU-Hmass}.

Our strategy to evaluate the Higgs boson mass is
 practically the same as the calculation of
 the instability bound on the Higgs boson mass
 if we identify the compactification scale
 with a cutoff for the SM.
However, considering that
 the gauge and Yukawa interactions are unified
 in the gauge-Higgs unification scenario,
 we naturally determine the compactification scale
 at which the running SU(2) gauge and top quark Yukawa
 couplings are unified.
As we will see later, such a compactification scale
 is found to be around $10^9$ GeV,
 and therefore the prediction of the gauge-Higgs unification
 scenario for the Higgs boson mass is compatible
 with the observed Higgs boson mass of around 125 GeV.

We first give a brief review of the gauge-Higgs unification scenario.
Since our purpose is to show the basic structure of
 the gauge Higgs unification scenario,
 we only consider a simple toy model based on
 an SU(3) gauge group with one SU(3) triplet bulk fermion ($\Psi$).
The fifth dimension is compactified on the $S^1/Z_2$ orbifold.
The model Lagrangian is expressed as
\bea
{\cal L} = -\frac{1}{2} \mbox{Tr}  (F_{MN}F^{MN})
+ i\bar{\Psi} \Gamma^M D_M \Psi,
\label{lagrangian}
\eea
where $M,N = 0,1,2,3,5$,
 the gamma matrix in 5-dimensional theory is defined as
 $\Gamma^M=(\gamma^\mu, i \gamma^5)$,
 $F_{MN}=\partial_M A_N - \partial_N A_M -i g_{5} \left[A_M, A_N
\right]$ with $ A_{M} = A_{M}^{a} \frac{\lambda^{a}}{2} $
 ($\lambda^{a}$: Gell-Mann matrices),
 the covariant derivative $D_M = \partial_M -ig_{5} A_M$,
 and $\Psi = (\psi_1 ~\psi_2 ~\psi_3)^T$.
While the periodic boundary condition along $S^1$ is imposed
 for all fields, non-trivial $Z_2$ parities are assigned
 for each field using the orbifolding matrix $P={\rm diag}(-,-,+)$
 such as
\bea
\label{z2parity}
A_\mu(-y) = P A_\mu(y) P^{-1}, \quad
A_5(-y) = -P A_5(y) P^{-1}, \quad
\Psi(-y) =  P \gamma^5 \Psi(y),
\eea
where $y$ is the fifth coordinate,
 and $\pm$ means even/odd $Z_2$ parities.
With these boundary conditions, the SU(3) gauge symmetry
 is broken to the electroweak gauge group SU(2)$\times$U(1).

The zero mode of the SU(3) gauge bosons is decomposed into
 the electroweak gauge boson of the SM
 and the Higgs doublet such that
\bea
A^{(0)}_{\mu} = \frac{1}{2}
\left(
\begin{array}{ccc}
 W^{3}_{\mu}+ \frac{1}{\sqrt{3}} B_{\mu} & \sqrt{2} W^{+}_{\mu} & 0 \\
\sqrt{2} W^{-}_{\mu} & - W^{3}_{\mu}+ \frac{1}{\sqrt{3}} B_{\mu}& 0 \\
 0& 0 & -\frac{2}{\sqrt{3}} B_{\mu}
\end{array}
\right), ~~~
A_5^{(0)} = \frac{1}{\sqrt{2}}
\left(
\begin{array}{ccc}
0 & 0 & h^{+} \\
0 & 0 & h^{0} \\
h^{-} & h^{0\ast} & 0
\end{array}
\right),
\eea
where $W_\mu^{3}, \ W_{\mu}^{\pm}$ and $B_\mu$
 are respectively the SU(2) and U(1) gauge bosons
 of the SM and $h = (h^{+}, h^{0})^{T}$
 is the Higgs doublet field\footnote{
Note that the toy model predicts an unrealistic
 weak mixing angle, $\sin^{2} \theta_{W} = \frac{3}{4}$.
This problem is ameliorated by introducing an additional U(1)
 gauge interaction and extending the toy model to
 a model based on SU(3)$\times$U(1)$^\prime$~\cite{GHU-Models},
 realizing the $Z$ boson as a combination of $B_\mu$
 and the extra U(1)$^\prime$ gauge boson. }.
Substituting  the zero mode expressions in the first term
 in Eq.~(\ref{lagrangian}), we obtain the kinematic terms
 for the SU(2)$\times$U(1) gauge bosons and
 the Higgs doublet with the correct covariant derivative,
 but no Higgs potential.

The zero mode of the bulk fermion is decomposed into
 an SU(2) doublet left-handed fermion and
 an SU(2) singlet right-handed fermion.
Their U(1) charges are the same as those
 of the quark SU(2) doublet and the SU(2) singlet
 down-type quark in this toy model\footnote{
In the SU(3)$\times$U(1)$^\prime$ extension,
 the fermion zero mode can be identified
 with the top quark by a suitable U(1)$^\prime$ charge assignment.
}.
The gauge and Yukawa interactions of the fermion
 originate from the higher dimensional gauge interaction
 and as a result, the SU(2) gauge coupling constant
 is the same as the Yukawa coupling constant
 at the compactification scale.
This relation may be suitable for the weak boson and
 top quark masses, since their masses are of the same order
 of magnitude.
As we will see below, the SU(2) gauge and top Yukawa couplings
 are unified at an intermediate scale and this scale
 is naturally identified with the compactification scale \cite{GHU-Hmass}.

\begin{figure}[ht]
  \begin{center}
   \includegraphics[width=75mm]{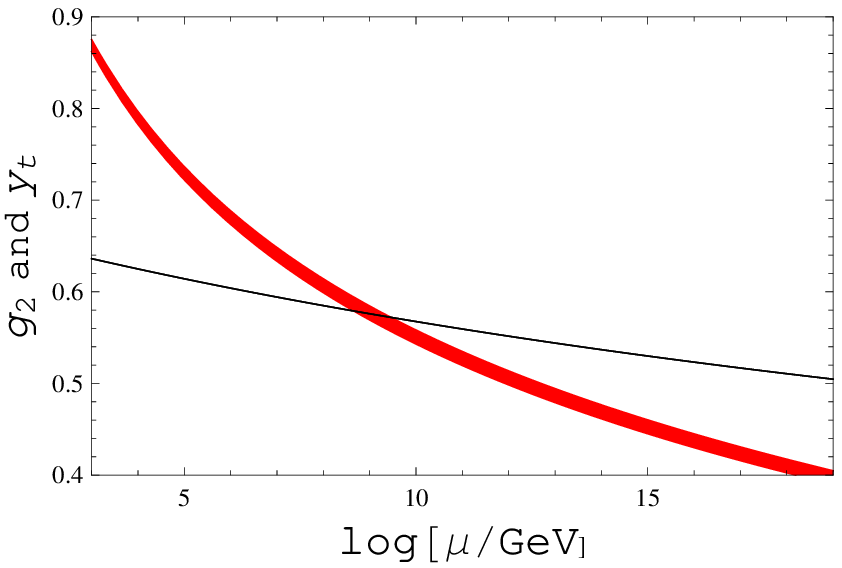}
   \hspace*{10mm}
   \includegraphics[width=78mm]{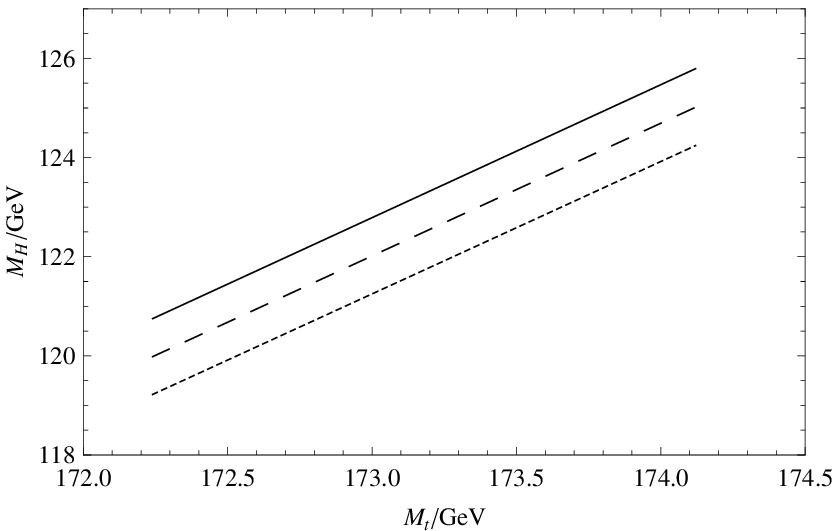}
   \end{center}
  \caption{
({\bf Left panel})
Renormalization group running of the SU(2) gauge coupling
 (shaded in black) and top Yukawa coupling (shaded in red)
 for various values of inputs in the range of
 $0.1179 \leq \alpha_3(m_Z) \leq 0.1193$
 and $172.24~{\rm GeV} \leq M_t \leq 174.12~{\rm GeV}$.
They unify at $\mu \simeq 10^9$ GeV,
 which is identified with the compactification scale.
({\bf Right panel})
Predicted Higgs boson mass in the gauge-Higgs unification
 scenario for various values of the top quark pole mass
 ($172.24 \leq M_t/{\rm GeV} \leq 174.12$)
 and three different values of the QCD coupling constant,
 $\alpha_3(m_Z)=0.1179$ (solid line),
 $0.1186$ (dashed line) and $0.1193$ (dotted line).
}
  \label{diphotonwithleptonR}
\end{figure}

In our analysis we employ the renormalization group
 equations of the SM at the two loop level~\cite{RGE}.
Since the resultant Higgs boson mass is sensitive to
 the top quark pole mass ($M_t$)and the QCD coupling constant
 $\alpha_{3}(M_{Z})$ at the $Z$ boson mass scale,
 we take into account experimental uncertainties
 in measuring these values:
 $M_t=173.18 \pm 0.94$~\cite{Tevatron} and
 $\alpha_{3}(m_Z)=0.1186 \pm 0.0007$~\cite{alpha_s}.
For the SU(2) and U(1) gauge couplings of the SM,
 we have used $\alpha_1(m_Z)=0.01618$ and $\alpha_2(m_Z)=0.03354$.
For various values of inputs
 in the range of
 $172.24 \leq M_t/{\rm GeV} \leq 174.12$ and
 $0.1179 \leq \alpha_3(m_Z) \leq 0.1193$,
 the running top Yukawa coupling (shaded in red)
 and SU(2) gauge coupling (shaded in black)
 are depicted in Fig.~1 (left panel).
The two couplings unify at $\mu \simeq 10^9$ GeV.
As $M_t$ is raised ($\alpha_3(m_Z)$ is lowered),
 the unification scale becomes slightly larger.

For fixed values of $M_t$ and $\alpha_3(m_Z)$,
 we first determine the compactification scale
 as the merger point of the running top Yukawa
 coupling and SU(2) gauge coupling.
Imposing the gauge-Higgs condition at this compactification scale,
 we run the Higgs quartic coupling down to low energies.
The pole mass of the Higgs boson is evaluated
 as a solution to the matching condition
 between the running mass and physical mass~\cite{MhPole}.
For various values of $M_t$ and $\alpha_3(m_Z)$,
 the Higgs boson mass is shown in Fig.~1 (right panel).
Three lines from top to bottom corresponds to
 input values of QCD coupling as
 $\alpha_3(m_Z)=0.1179$ (solid), $0.1186$ (dashed)
 and $0.1193$ (dotted), respectively.
As shown in the figure,
 the Higgs boson mass prediction by the gauge-Higgs unification
 scenario is compatible with the measured Higgs boson mass
 of around 125 GeV within the experimental uncertainties
 in measurements of the top quark pole mass and the QCD coupling constant.
Larger (smaller) values for $M_t$ ($\alpha_3(m_Z)$)
 are preferable.
More precise measurement of the Higgs boson mass
 along with top quark pole mass and the QCD coupling constant
 is therefore crucial for testing the Higgs boson mass prediction
 by the gauge-Higgs unification scenario.

\section*{Acknowledgments}
This work is supported in part by the DOE Grants,
  \# DE-FG02-12ER41808 (I.G. and Q.S.),  \# DE-FG02-10ER41714 (N.O.)
   and by Rustaveli National Science Foundation No.~31/89 (I.G.).
I.G. would like to thank CETUP* (Center for Theoretical Underground 
Physics and Related Areas), supported by the US Department of Energy 
under Grant No. DE-SC0010137 and by the US National Science Foundation 
under Grant No. PHY-1342611, for its hospitality and partial support 
during the 2013 Summer Program


\end{document}